\documentstyle[twocolumn,aps,graphicx,prb]{revtex}
\begin{document}
\draft
\title{A magnetic barrier in confined two-dimensional electron gases:
  Nanomagnetometers and magnetic switches}
\author {Michele Governale} 
\address{Institut f\"ur Theoretische Festk\"orperphysik, Universit\"at
  Karlsruhe, D-76128 Karlsruhe, Germany,\\ and 
Dipartimento di Ingegneria dell'Informazione,
Universit\`a di Pisa, 
Via Diotisalvi, 2, I-56126 Pisa, Italy\\} 
\author{Daniel Boese}
\address{Institut f\"ur Theoretische Festk\"orperphysik, Universit\"at
  Karlsruhe, D-76128 Karlsruhe, Germany,\\ and
Forschungszentrum Karlsruhe, Institut f\"ur Nanotechnologie, D-76021
Karlsruhe, Germany}
\date{\today} 
\maketitle
\begin{abstract}
We investigate the conductance properties of a hybrid
ferromagnet-semiconductor structure consisting of a confined two-dimensional
electron gas and a transverse ferromagnetic strip on top. Within the framework
of the Landauer-B\"uttiker model, we develop an alternative way to consider 
magnetic
fields. Our method describes devices ranging from a recently
realized nanomagnetometer down to quasi one-dimensional quantum wires. 
We provide a 
rigorous way to 
 relate the measured resistance to the actual magnetization
of the strip. Regarding the quasi one-dimensional wires we propose
a new device application, a tunable magnetic switch. 
\end{abstract}
\pacs{PACS numbers: 73.34.Ad, 72.15.Gd  }
\narrowtext
Hybrid ferromagnet-semiconductor structures recently have attracted much
attention.\cite{Prinz,Kent,Ye,Johnson,Reijniers,Geim} Such a combination for
instance would allow to probe magnetic 
properties on a nanometer scale, or to build devices such as
spin-valves or
nanomagnetometers\cite{Kubrak}. In this letter we are
interested in the latter class of structures, where
the spin of the electron plays a minor role. 
After a brief review of the experimental situation we present the model and
the method used 
to describe the full quantum mechanical problem. We then discuss the results
obtained and compare them with the experiment. A possible device application
is discussed at the end.

\begin{figure}
       \centerline{\includegraphics[width=7cm]{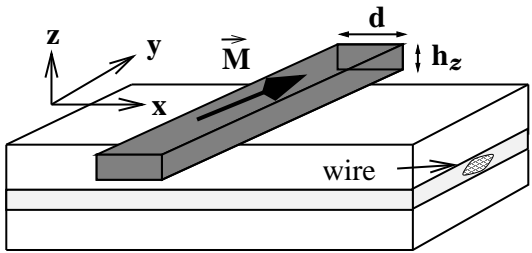}}
       \vspace{0.5cm}
                 \caption{Geometry of the system. The magnetization $\mathbf M$ 
                  of the
                   strip can be rotated by applying a field in the
                  $x$-direction. 
                  }
         \protect\label{fig:geo}
\end{figure}

\begin{figure}
       \centerline{\includegraphics[width=7cm]{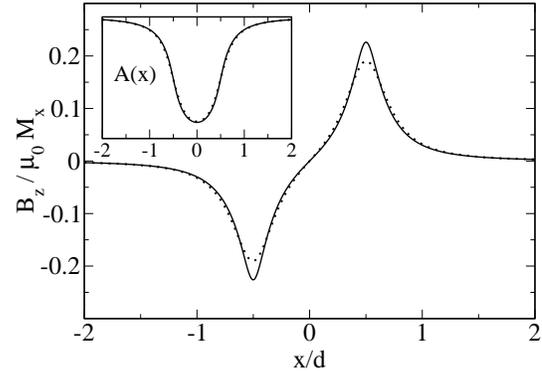}} \vspace{0.5cm}
                 \caption{Shape of $B_z (x)$ and the corresponding vector
                   potential $A_y(x)$ in the inset. The solid line is the
                  exact solution, the dotted one shows the approximation for
                  $h_z \ll z_0$. 
                  }
         \protect\label{fig:hzshape}
\end{figure}

A device that has been experimentally realized\cite{Kubrak} consists of a
two-dimensional electron gas (2DEG) which flows through a lateral
constriction, i.e. a quantum wire, with a
ferromagnetic strip placed on top of it (see Fig.~\ref{fig:geo}). The
linear conductance along the longitudinal direction acts as a probe for
magnetic fields. As the $g$-factor for the 2DEG is relatively small the spin
degree of freedom enters only as a factor of $2$ in the conductance.
Only the perpendicular component of the magnetic field $B_z$ can affect the
transport properties of the system. However an external magnetic field
$H_{\mathrm{ext}}$ in the longitudinal direction can be 
applied in order to rotate the strip's magnetization $\mathbf{M}$. For
$H_{\mathrm{ext}}=0$, $\mathbf{M}$ is assumed to point into the transverse
direction, due to symmetry this results in zero perpendicular magnetic field
$B_z$.  
As soon as $\mathbf{M}$ acquires a non-zero $x$-component, a finite $B_z$ 
appears which acts as a magnetic barrier in the wire. A typical profile of
$B_z$ is shown in Fig. \ref{fig:hzshape}. 
It is obvious that the the presence of such a barrier strongly alters the
conductance properties. Kubrak {\em{et al.}}\cite{Kubrak} measured the change
in the longitudinal conductance as a function 
of $H_{\mathrm{ext}}$. They observed a decrease in the conductance with
increasing $|H_{\mathrm{ext}}|$ until it saturates at a finite value. Moreover
they could measure a hysteresis, from which one can conclude that the saturation
is reached when $\mathbf{M}$ points entirely into the longitudinal
direction\cite{note1}.

In order to explain their experimental findings Kubrak {\em{et al.}} applied a
semi-classical theory developed for periodic structures. Although some
comparison could be done, we think a more rigorous calculation
should be performed. Moreover in our model we take into account the transverse
confinement and this allows us to go from a quasi 1D quantum wire (few
propagating channels) to the Hall-bar situation of the experiment (many
propagating channels). The work presented in this letter is therefore
more applicable than full 2D theories\cite{Matulis} and describes more realistic
electro-magnetic fields than previous quantum wire model
calculations.\cite{Takagaki}  

Our approach is based on the  Landauer-B\"uttiker model which is appropriate
if electron-electron interactions are negligible. The conductance is 
given by the sum over the transmission probabilities of all propagating
channels\cite{landbuett} 
\begin{equation}
G=\frac{2 e^2}{h} \sum_{n,m=1}^N T_{nm} .
\end{equation} 
Usually the magnetic field is introduced by choosing a gauge in which the
vector potential has only a non-vanishing component along the direction of
current flow.\cite{Takagaki,Schult} This leads to a dependence of the 
transverse wave function on the 
longitudinal $k$-vectors, which in addition are complex. On the contrary we
choose a gauge where the vector potential is all along the transverse
direction:
\begin{equation}
{\mathbf{A}}(x,y)={\mathbf{A}}(x) =\int \!\! dx \, B_z(x) \, \,
{\mathbf{\hat{y}}}. 
\end{equation}
We now introduce a discretization grid $\{ x_i \}$ for the longitudinal
direction. Thus we cut the wire into thin slices bounded by the grid points and
assign a constant vector potential to each slice. In every slice the
Hamiltonian can be decomposed in a longitudinal (free motion) and a transverse
part, that contains all the information on the magnetic field: 
\begin{eqnarray}
H^{\mathrm{long}}&=& - \frac{\hbar^2}{2m} \partial_x^2 \nonumber \\
H_{x_i}^{\mathrm{trans}}&=& \frac{1}{2m} \left( -i \hbar \partial_y + e A_{x_i}
\right)^2 + V_{x_i}(y) ,  
\end{eqnarray} 
where $A_{x_i}=A(x_i)$ and $V_{x_i}(y)$ is the transverse confining
potential in the $i$-th slice.  
The presence of the gauge field changes the transverse solution in every slice
like  
\begin{eqnarray}
\chi_{n,x_i}(y) &=& \chi_n^0(y) \, \, e^{-i e/\hbar A_{x_i} y} \nonumber \\
E_{n,x_i}&=&E_{n,x_i}^0, 
\end{eqnarray}
where quantities with a $0$ superscript refer to zero magnetic field.
Hence the vector potential leaves the transverse eigenenergies unchanged and 
manifests itself only in a local phase of the wave function. Finally we
use a scattering matrix method to compute the transmission through the entire
structure.\cite{Datta} One big advantage of our method is that an increase in 
complexity of the device' or barrier's structure does not affect the
simulation setup, due to its modular architecture. 

In the present work we consider a hard wall confining potential of 
width $W$. 
The transport problem is fully described by the transverse eigenenergies 
in each slice $E_{n,x_i}$, and the overlap integrals between transverse wave 
functions belonging to neighboring slices
$S_{nm}=\left\langle \chi_{n,x_i} | \chi_{m,x_{i+1}} \right\rangle$. 
The vector potential appears in the 
expressions for the overlap integrals only as 
$eW/\hbar \left( A_{x_{i+1}} - A_{x_i} \right) =eW/\hbar\int_{x_i}^{x_{i+1}}
\!\!dx' \, B(x')= \Phi_{x_i}/\Phi'$, where $\Phi_{x_i}$  
is the magnetic flux in the slice that goes from $x_i$ to $x_{i+1}$, 
and $\Phi'$ is the flux quantum. 
The previous remark assures us that the method is not affected by the 
problems that may arise in discretizing a gauge field, as the results 
manifestly depend only on the spatial distribution of magnetic flux. 
The overlap integrals also contain the information of how much mode-mixing 
occurs at the interfaces between the slices; we use this property to 
obtain a criterion for selecting the longitudinal discretization mesh. 
In particular, we require magnetic induced mode-mixing to be small, and
this condition  
translates into $\Phi_{x_i}/\Phi' < 1$. Thus the discretization grid has to be
chosen  
so that each transverse slice contains less than a flux quantum. 
In our numerical calculation this is assured
by an adaptive procedure for selecting the longitudinal mesh.\cite{numslice}
The magnetic field used for the calculations is computed from the equivalent
surface pole 
densities of $M_x$ \cite{Craik,Kubrak} with the approximation that the
height of the strip $h_z$ is 
much smaller than the distance of the 2DEG from the strip's center $z_0$. 
The width of the strip shall be denoted as $d$.
In Fig.~\ref{fig:hzshape} we show the magnetic field profile, computed making 
use of the afore mentioned approximation and without it. Although there is a 
small deviation for the peak values, the approximation is a good one 
even for strip heights that are of the same order of $z_0$. 
  
\begin{figure}
       \centerline{\includegraphics[width=7cm]{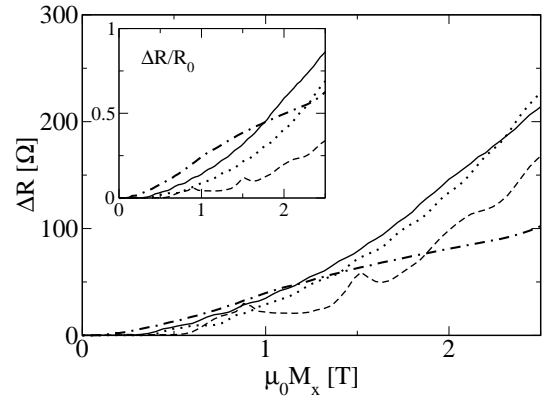}}\vspace{0.5cm}
                 \caption{Longitudinal resistance change vs longitudinal
                  magnetization for different wire width $W=1500 
                  {\mathrm nm}$ (dot-dashed), $W=1000 {\mathrm nm}$
                  (solid), $W=750 {\mathrm nm}$ (dotted), 
                  and $W=500 {\mathrm nm}$ (dashed). Parameter
                  values are $E_F=15.7 {\mathrm meV}$, $d=400 {\mathrm nm}$,
                  $h_z=120 {\mathrm nm}$, $z_0=95 {\mathrm nm}$. Inset:
                  Longitudinal resistance change normalized to the zero
                  field value for the same parameter values.
                  }
         \protect\label{fig:RvsMwide}
\end{figure}

The first results we show regard a wide wire and its potential application as 
a nanomagnetometer.  
In Fig.~\ref{fig:RvsMwide} we show the longitudinal resistance 
change $\Delta R$ 
of a two-point measurement as a function of the longitudinal magnetization. 
This plot
allows to extract $M_x$ from the measured $\Delta R$, which then can be used  
to derive the $M_x$ vs $H_{\mathrm ext}$ characteristics (hysteresis curve) for 
the real experimental situation. The solid line corresponds to the experimental
setup.\cite{Kubrak}
The inset shows the sensitivity of the resistance to the magnetic field. 
Decreasing the width of the
wire leads to lower values of the sensitivity, due to the fact 
that the wire picks up less magnetic flux. 
In addition little bumps appear, caused by the increasing influence of 
the lateral confinement. 
The computed $\Delta R$-$M_x$ relation is in good agreement with the 
experimental findings. 
 
\begin{figure}
       \centerline{\includegraphics[width=7cm]{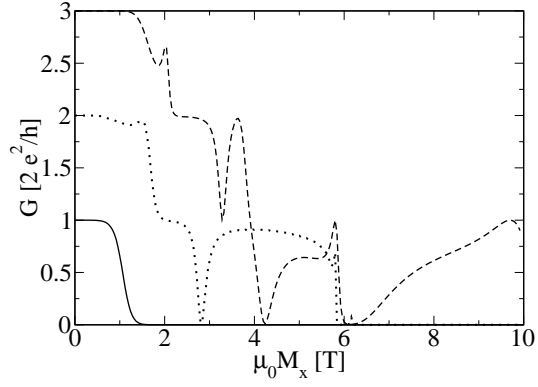}}\vspace{0.5cm}
                 \caption{Conductance vs longitudinal
                  magnetization in a narrow wire ($W=200 {\mathrm nm}$ for
                  different Fermi Energies $E_F= 1.5 E_1$ (solid), $E_F= 6.5
                  E_1$ (dotted) and $E_F= 12.5 E_1$ (dashed).  Barrier
                  parameters: $d=400 {\mathrm nm}$, 
                  $h_z=120 {\mathrm nm}$, $z_0=95 {\mathrm nm}$.
                  }
         \protect\label{fig:GvsMnarrow}
\end{figure}

From now on we will present results for the case of narrow wires, where 
the quantization along the transverse direction becomes crucial. 
In Fig.~\ref{fig:GvsEfnarrow} we show the linear conductance vs the Fermi
Energy. For small $M_x$ the conductance resembles
the zero field case, but it shows little oscillations around channel
openings. These oscillations develop into well-defined resonances, that only
manifest themselves in the highest propagating mode. For large 
magnetization values one clearly can identify the shift  
of the conductance plateaus\cite{Takagaki} towards higher energies. 
This depletion of the conducting channels by the magnetic field becomes more
obvious in Fig.~\ref{fig:GvsMnarrow}, where  we plot the conductance as a
function of the magnetization for 
three different Fermi energies (indicated by arrows in
Fig.~\ref{fig:GvsEfnarrow}). Although we have increased the magnetization 
to values that are higher than experimentally achievable with ferromagnetic 
materials (e.g. the saturation 
value for cobalt is $\mu_0 M_s\approx 1.8$T), this is interesting not only 
for theoretical reasons, but also because high-field situations could be 
realized by different experimental setups. 
However we wish to point out that the abrupt 
drop of the conductance for the one mode case occurs when the magnetization 
exceeds an experimentally accessible threshold value. 
 
In virtue of the previous remark we propose that this device, in 
the few conducting channel regime, can be used as a 
{\em tunable magnetic switch}. 
The switch is open if the
component of the external magnetic field along the longitudinal direction is
large enough to  rotate the magnetization $\mathbf M$ 
such that it magnetically depletes the wire. 
The field value at which the switching occurs can be tuned 
by varying the Fermi Energy (or the width) of the wire, or the geometrical 
parameters of the ferromagnetic strip. 
We would like to remark that all the key building blocks for 
such a device (shallow 2DEG, few modes quantum wire\cite{Liang} and 
ferromagnetic strip deposition) have been experimentally realized.    

In conclusion we developed a method to describe transport through
a confined 2DEG with a magnetic barrier. We performed calculations which will
allow to relate the experimentally measured resistance to the properties of
the ferromagnetic strip (hysteresis curve). We also investigated 
much smaller systems. Based on our findings we propose to use these quantum
wires-ferromagnet hybrid structure as tunable magnetic switches.

The authors acknowledge financial support from the EU-Project ''Dynamics of 
nano-fabricated superconducting circuits'' (M.~G.) and the DFG
Graduiertenkolleg ''Kollektive Ph\"anomene im Festk\"orper'' (D.~B.).

\begin{figure}
       \centerline{\includegraphics[width=7cm]{fig4}}\vspace{0.5cm}
                 \caption{Conductance vs Fermi Energy (in units of the first
                   subband energy $E_1$) in a narrow wire 
                  $W=200 {\mathrm nm}$ for different magnetization $\mu_0 
                  M_x= 0.1{\mathrm T}$ (solid),    
                  $\mu_0 M_x= 0.5 {\mathrm T}$ (dashed),
                  $\mu_0 M_x= 1.0 {\mathrm T}$ (dotted), $\mu_0 M_x= 2.0 
                  {\mathrm T}$
                  (long dashed) and  $\mu M_0= 5.0 {\mathrm T}$ (dot-dashed).
                  Barrier parameters: $d=400 {\mathrm nm}$,
                  $h_z=120 {\mathrm nm}$, $z_0=95 {\mathrm nm}$. 
                  }  
         \protect\label{fig:GvsEfnarrow}
\end{figure}

\begin{references}
\bibitem{Prinz} G.~A.~Prinz, Science {\bf 250}, 1092 (1990), {\em ibid.} {\bf
    282}, 1660 (1998). 
\bibitem{Kent}A.~D.~Kent, S.~von Moln\'ar, S.~Gider, and D.~D.~Awschalom,
  J. Appl. Phys. {\bf{76}}, 6656 (1994).
\bibitem{Ye}P.~D.~Ye, D.~Weiss, R.~R.~Gebhardts, M.~Seeger, K.~von~Klitzing,
  K.~Eberl, and H.~Nickel, Phys. Rev. Lett. {\bf{74}}, 3013 (1995).
\bibitem{Johnson} M.~Johnson, B.~R.~Bennett, M.~J.~Yang, M.~M.~Miller, and
  B.~V.~Shanabrook, Appl. Phys. Lett. {\bf{71}}, 974 (1997).
\bibitem{Reijniers}J.~Reijniers and F.~M.~Peeters,
  Appl. Phys. Lett. {\bf{73}}, 357 (1998).
\bibitem{Geim}A.~K.~Geim, S.~V.~Dubonos, J.~G.~S.~Lok, I.~V.~Lok,
  I.~V.~Grigorieva, J.~C.~Maan, L.~Theil~Hansen, and P.~E.~Lindelof,
  Appl. Phys. Lett. {\bf{71}}, 2379 (1997). 
\bibitem{Kubrak}
V. Kubrak, F.~Rahman, B.~L.~Gallagher, P.~C.~Main, M.~Henini, C.~H.~Marrows, and 
M.~A.~Howson, Appl. Phys. Lett. {\bf{74}}, 2507 (1999).
\bibitem{note1} The magnetization $\mu_0\mathbf M$ in  the present work 
corresponds to $\mathbf J$ of Ref.~\onlinecite{Kubrak}.
\bibitem{Matulis}
A. Matulis, F.~M. Peeters, P. Vasilopoulos, Phys. Rev. Lett {\bf{72}}, 1518
(1994).
\bibitem{Takagaki}
Y. Takagaki, K. Ploog, Phys. Rev. B {\bf{51}}, 7017 (1995).
\bibitem{landbuett}
R. Landauer, IBM J. Res. Dev. {\bf{1}}, 223 (1957), M. B\"uttiker, IBM
J. Res. Dev. {\bf{32}}, 306 (1988).
\bibitem{Schult}
R.~L.~Schult, H.~W.~Wyld and D.~G.~Ravenhall, Phys. Rev. B {\bf 41}, 12760
(1990).
\bibitem{Datta}
M. Cahay, M. McLennan, S. Datta, Phys. Rev. B {\bf{37}}, 10125 (1988).
\bibitem{numslice}
In this work we consider up to 1000 slices and 150 modes.
\bibitem{Craik} For an outline of the computation see, e.g., 
D.~Craik, {\em Magnetism: Principles and Applications} (Wiley,
  Chicester, 1995), Chap. 4.1.17.
\bibitem{Liang}C.~T. Liang, M.~Y.~Simmons,
  C.~G.~Smith, D.~A.~Ritchie and M.~Pepper, Appl. Phys. Lett. {\bf 75}, 2975
  (1999).
\end{references}
\end{document}